\documentclass{mem}
\usepackage{natbib}\usepackage{txfonts}\usepackage{balance}
\usepackage{graphicx}
\usepackage[a4paper]{hyperref}
\idline{1}{1}
\begin{document}

\def\lsun{L$_\odot$}
\def\msun{M$_\odot$}
\def\mstar{$M_\star$}
\def\mgas{$M_{gas}$}
\def\fgas{$f_{gas}$}
\def\zsun{Z$_\odot$}
\def\mc{\multicolumn}
\def\cgs{erg~cm$^{-2}$sec$^{-1}$}
\def\mic{$\mu$m}
\def\arcsec{$^{\prime\prime}$}
\def\yeff{$y_{eff}$}
\def\ysun{$y_\odot$}

\def\ha{H$\alpha$}
\def\hb{H$\beta$}
\def\oii{[OII]$\lambda$3727}
\def\nii{[NII]$\lambda$6584}
\def\neiii{[NeIII]$\lambda$3869}
\def\oiiia{[OIII]$\lambda$4958}
\def\oiiib{[OIII]$\lambda$5007}
\def\oiii{[OIII]$\lambda$4958,5007}
\def\oiit{[OII]\small{3727}}
\def\neiiit{[NeIII]\small{3869}}
\def\oiiiat{[OIII]\small{4958}}
\def\oiiibt{[OIII]\small{5007}}
\def\hei{HeI\small{3889}}

\def\aj{AJ}
\def\apj{ApJ}
\def\apjs{ApJS}
\def\apjl{ApJL}
\def\aap{A\&A}	
\def\mnras{MNRAS}
\def\pasp{PASP}
\def\araa{ARAA}
\def\nat{Nature}

\title
{Galaxy metallicity near and far}

\author{
F. Mannucci\inst{1} \and
G. Cresci\inst{1,2}
}

\institute{
Istituto Nazionale di Astrofisica --
Osservatorio Astrofisico di Arcetri, 
Largo E. Fermi 5, I-50125, Firenze, Italy
\email{filippo@arcetri.astro.it}
\and
Max-Planck-Institut f\"ur extraterrestrische Physik (MPE), Giessenbachstr. 1, 85748, Garching, Germany
}

\authorrunning{Mannucci}

\titlerunning{Galaxy metallicity near and far}

\abstract{
Metallicity appears to be one the most important tool to study the formation
and evolution of galaxies. Recently, we have shown that
metallicity of local galaxies is tightly related not only to stellar mass, but also to 
star formation rate (SFR).
At low stellar mass, metallicity decreases sharply with increasing SFR,
while at high stellar mass, metallicity does not depend on SFR.
The residual metallicity dispersion across this 
Fundamental Metallicity Relation (FMR) 
is very small, about 0.05~dex.
High redshift galaxies, up to z$\sim$2.5, 
are found to follow the 
same FMR defined by local SDSS galaxies,
with no indication of evolution. 
At z$>$2.5, evolution of about 0.6~dex off the FMR is observed,
with high-redshift galaxies showing lower metallicities.
This result can be combined with our recent discover of metallicity gradients 
in three high redshift galaxies showing disk dynamics. In these galaxies, the
regions with higher SFR also show lower metallicities. Both these evidences
can be explained by the effect of smooth infall of gas into 
previously enriched galaxies, with the star-formation activity triggered 
by the infalling gas.
}
\maketitle

\section{Introduction}

Gas metallicity is regulated by a complex interplay between 
star formation, infall of metal-poor gas and outflow of enriched material. 
A fundamental discovery is the relation between stellar mass \mstar\ 
and metallicity \citep{McClure68,Lequeux79,Garnett02,Tremonti04,Lee06},
with more massive galaxies showing higher metallicities.
The origin of this relation is debated, and many different explanations
have been proposed, including ejection of metal-enriched gas
(e.g., \citealt{Edmunds90,Lehnert96a,Tremonti04}),
``downsizing'', i.e., a systematic dependence of the efficiency of star 
formation with galaxy mass  
(e.g., \citealt{Brooks07,Mouchine08,Calura09a}), variation of the IMF
with galaxy mass \citep{Koppen07}, and infall of metal-poor gas
\citep{Finlator08,Dave10}.

\begin{figure}
\centerline{\includegraphics[width=0.48\textwidth]{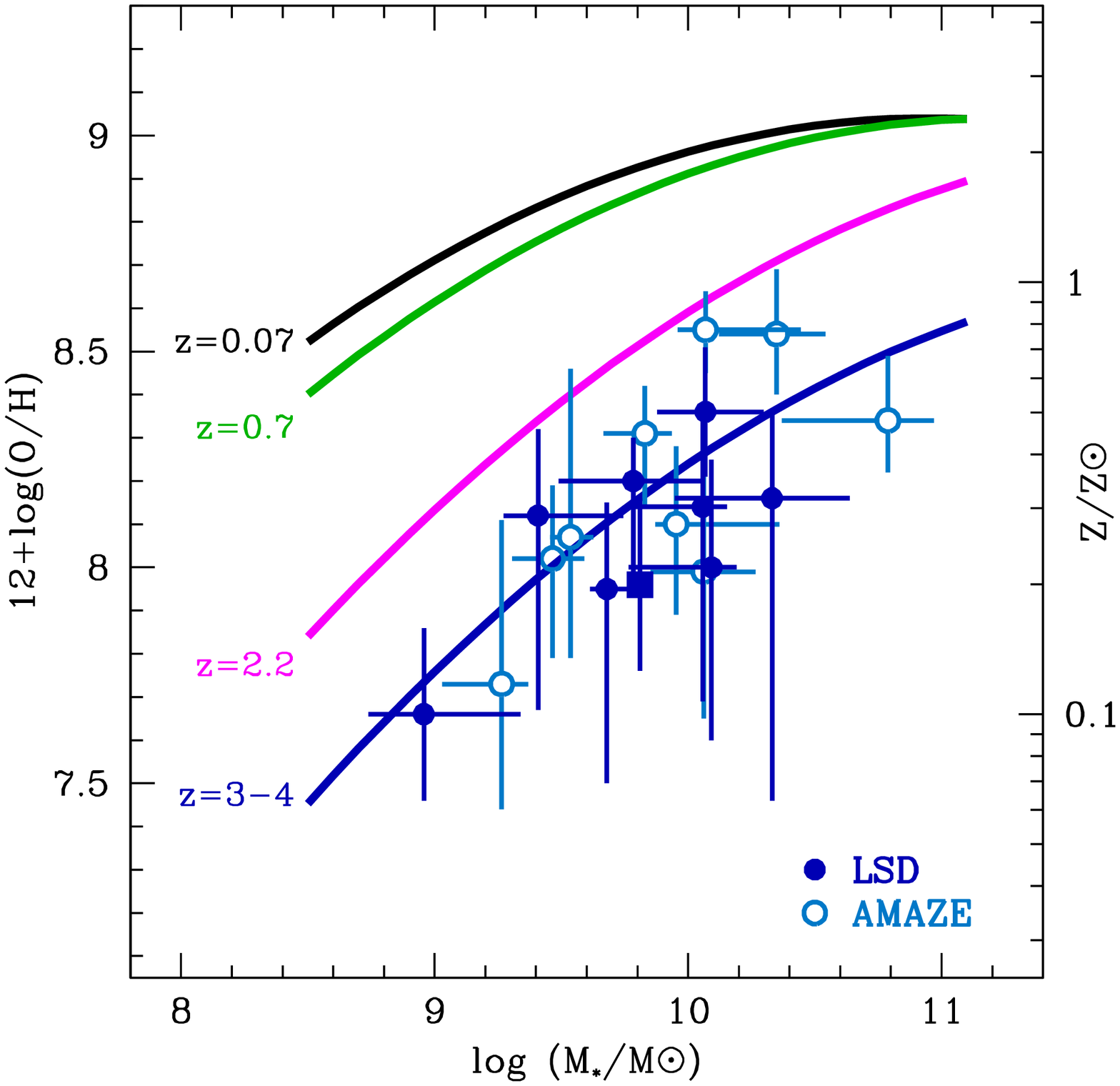} 
}
\caption{\footnotesize
Evolution of the mass-metallicity relation from local to high redshift galaxies
from \cite{Mannucci09b}.
Data are from \cite{Kewley08} (z=0.07), \cite{Savaglio05} (z=0.7), 
\cite{Erb06a} (z=2.2) and \cite{Mannucci09b} (z=3--4).
}
\label{fig:massmetevol}
\end{figure}

The mass-metallicity relation has been studied
by \cite{Erb06a}  at z$\sim$2.2 and by
\cite{Maiolino08} and \cite{Mannucci09b} at z=3--4, finding 
a strong and monotonic evolution, with metallicity decreasing with redshift
at a given mass (see fig.\ref{fig:massmetevol}.
The same authors \citep{Erb06a,Erb08,Mannucci09b} 
have also studied the relation between metallicity and gas fraction, i.e., 
the effective yields, obtaining clear 
evidence of the importance of infall in high redshift galaxies. 

If infall is at the origin of the star formation activity,
and outflows are produced by exploding supernovae (SNe),
a relation between metallicity and SFR
is likely to exist. In other words, SFR is a parameter
that should be considered in the scaling relations that include metallicity,
such as the mass-metallicity relation.
  
\begin{figure*}
\centerline{
   \includegraphics[width=0.48\textwidth]{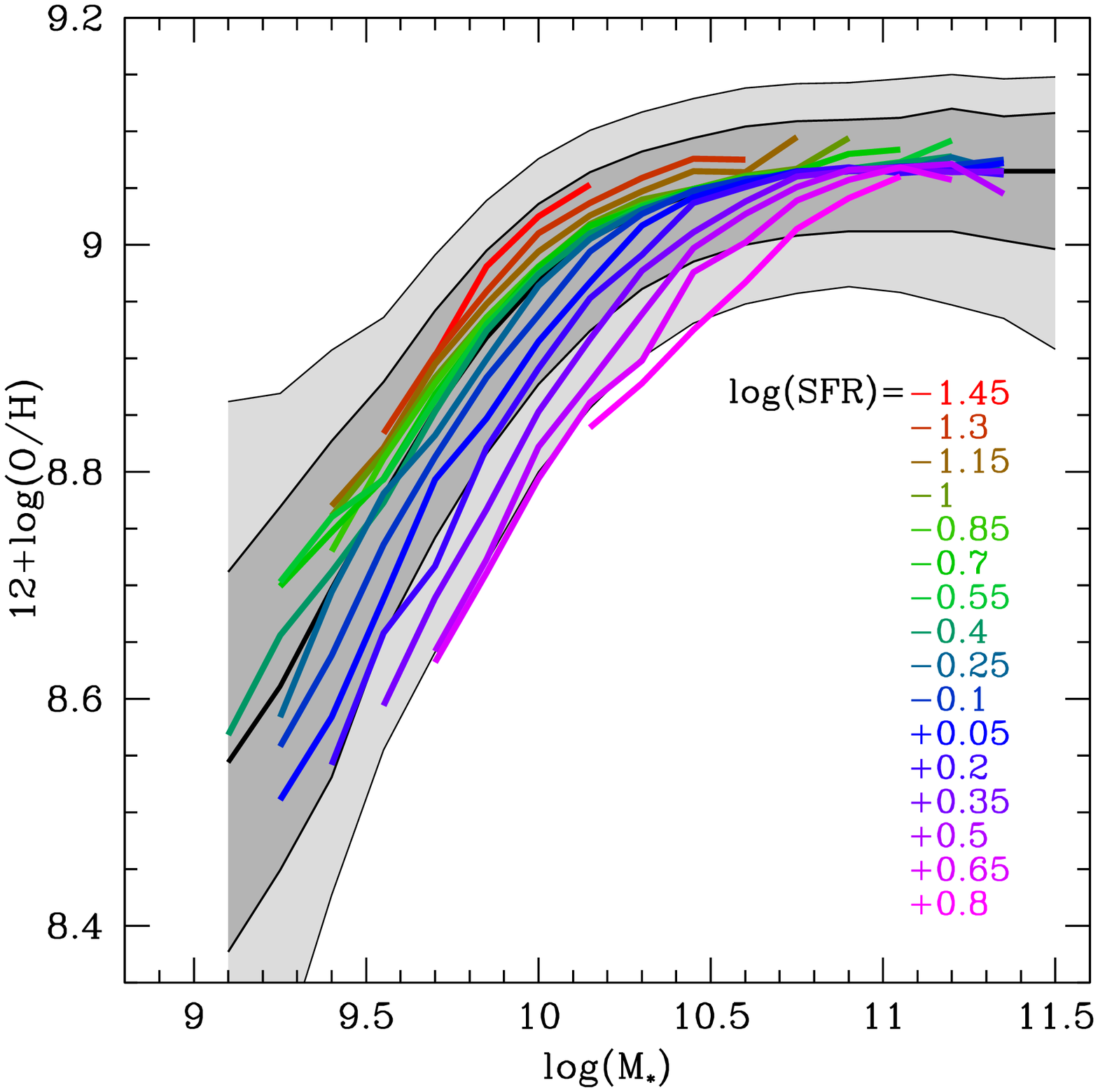} 
   \includegraphics[width=0.48\textwidth]{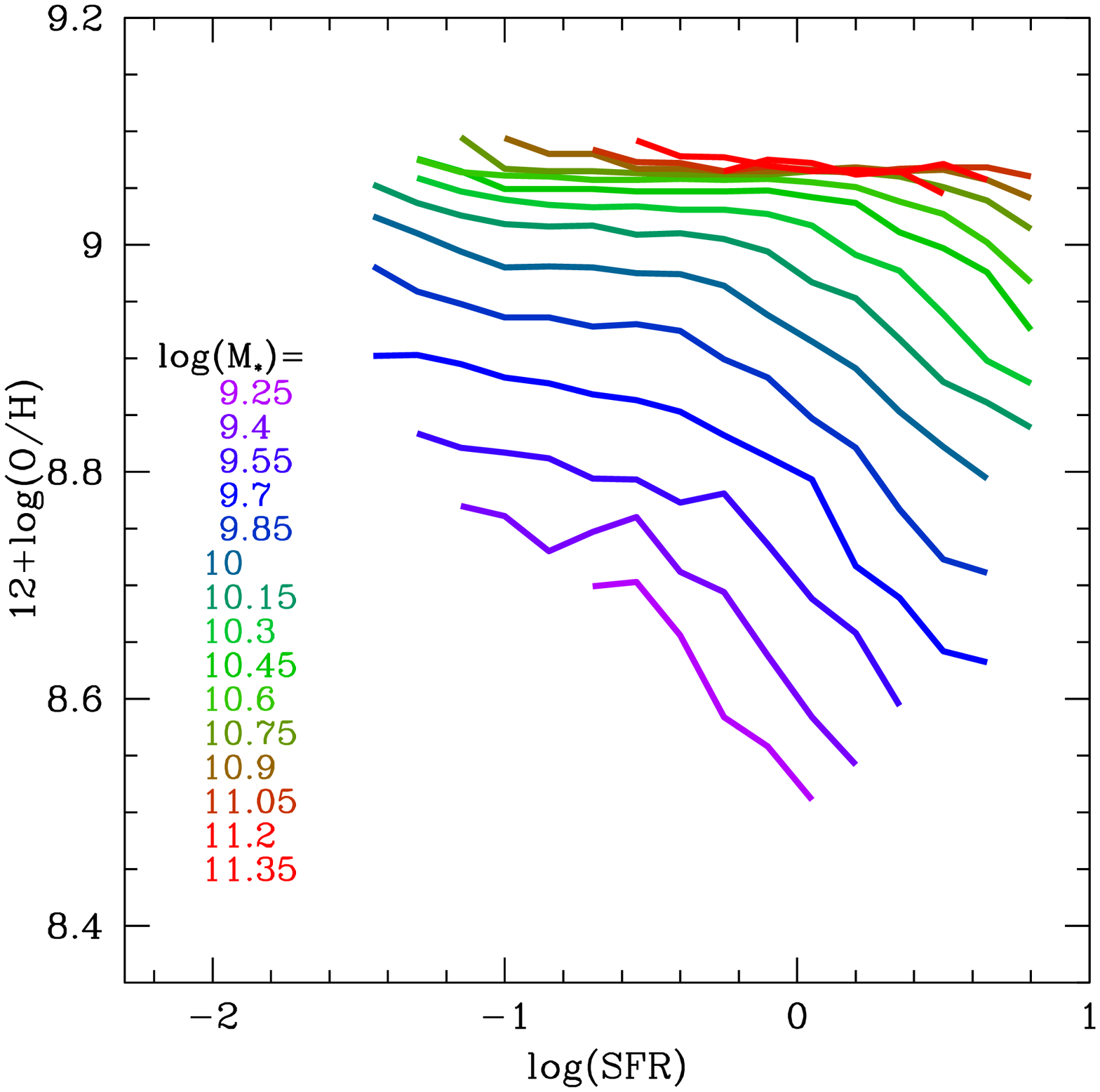} 
}
\caption{\footnotesize
{\em Left panel:} The mass-metallicity relation of local SDSS galaxies. 
The grey-shaded areas contain 64\% and 90\% of all SDSS galaxies, with the 
thick central line showing the median relation. The colored lines show the 
median metallicities, as a function of \mstar, 
of SDSS galaxies with different values of SFR. 
{\em Right panel:} median metallicity as a function of SFR for galaxies of 
different \mstar.
At all \mstar\ with log(\mstar)$<$10.7,
metallicity decreases with increasing SFR 
at constant mass .
}
\label{fig:massmet}
\end{figure*}

\section{The local Fundamental Metallicity Relation}

To test the hypothesis of a correlation between SFR and metallicity
in the present universe and at high redshift,
we have studied 
several samples of galaxies at different redshifts whose 
metallicity, \mstar, and SFR have been measured.
A full description of the data set is given in \cite{Mannucci10}

Local galaxies are well measured by the SDSS project
\citep{Abazajian09}. Among the $\sim10^6$ galaxies with observed spectra,
we selected star forming objects with redshift between 0.07 and 0.30,
having a signal-to-noise ratio (SNR) of \ha\ 
of SNR$>$25 and dust extinction $A_V<2.5$.
Total stellar masses \mstar\  from \cite{Kauffmann03a} were used,
scaled to the \cite{Chabrier03} initial mass function (IMF).
SFRs inside the spectroscopic aperture were measured 
from the \ha\ emission line flux corrected for dust extinction 
as estimated from the Balmer decrement.
The conversion factor between \ha\ luminosity and SFR 
in \cite{Kennicutt98} was used,
corrected to a \cite{Chabrier03} IMF.
Oxygen gas-phase abundances were measured from the emission line ratios
as described in \cite{Maiolino08}. An average
between the values obtain from \nii/\ha\ and 
R23=(\oii+\oiii)/\hb\ was used.
The final galaxy sample contains 141825 galaxies.

The grey-shaded area in the left panel of Fig.~\ref{fig:massmet} 
shows the mass-metallicity relation for our sample of SDSS galaxies. 
Despite the differences in the
selection of the sample and in the measure of metallicity, 
our results are very similar to what has been found by \cite{Tremonti04}.
The metallicity dispersion of our sample, $\sim$0.08~dex, is somewhat 
smaller to what
have been found by these authors, $\sim$0.10~dex, possibly due to different
sample selections and metallicity calibration.

The left panel of Fig.~\ref{fig:massmet} also shows, as a function of \mstar,
 the median metallicities
of SDSS galaxies having different levels of SFR. 
It is evident that a systematic segregation in SFR is present in the data.
While galaxies with high \mstar\ (log(\mstar)$>$10.9) show no correlation 
between metallicity and SFR, at low \mstar\ more active galaxies 
also show lower metallicity.
The same systematic dependence of metallicity on SFR can be seen in 
the right panel of Fig.~\ref{fig:massmet},
where metallicity is plotted as a function of SFR for different values of mass.
Galaxies with high SFRs show a sharp dependence of metallicity on SFR, while
less active galaxies show a less pronounced dependence.

\begin{figure*}[t]
\centerline{
	\includegraphics[width=0.32\textwidth]{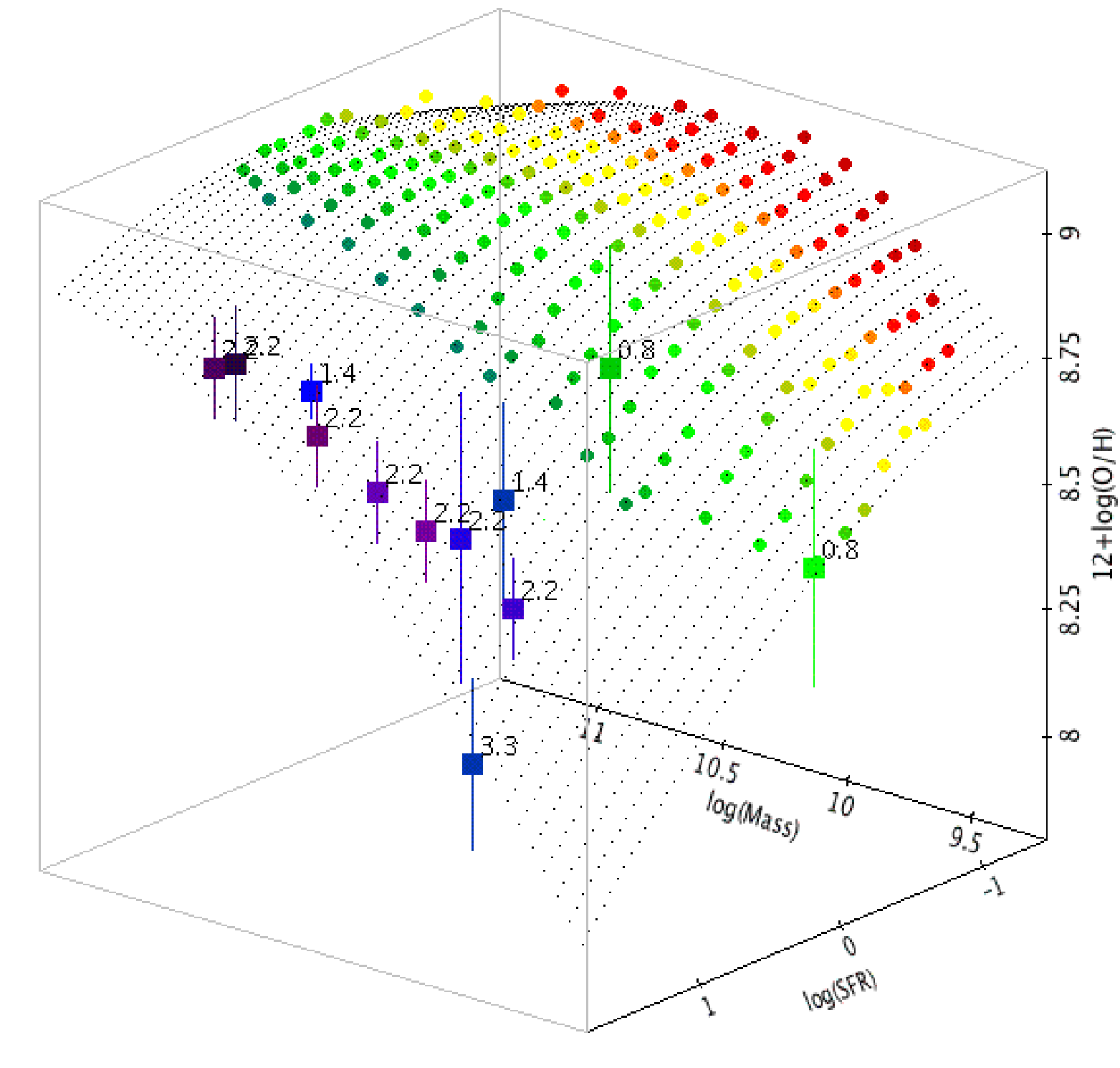} 
	\includegraphics[width=0.32\textwidth]{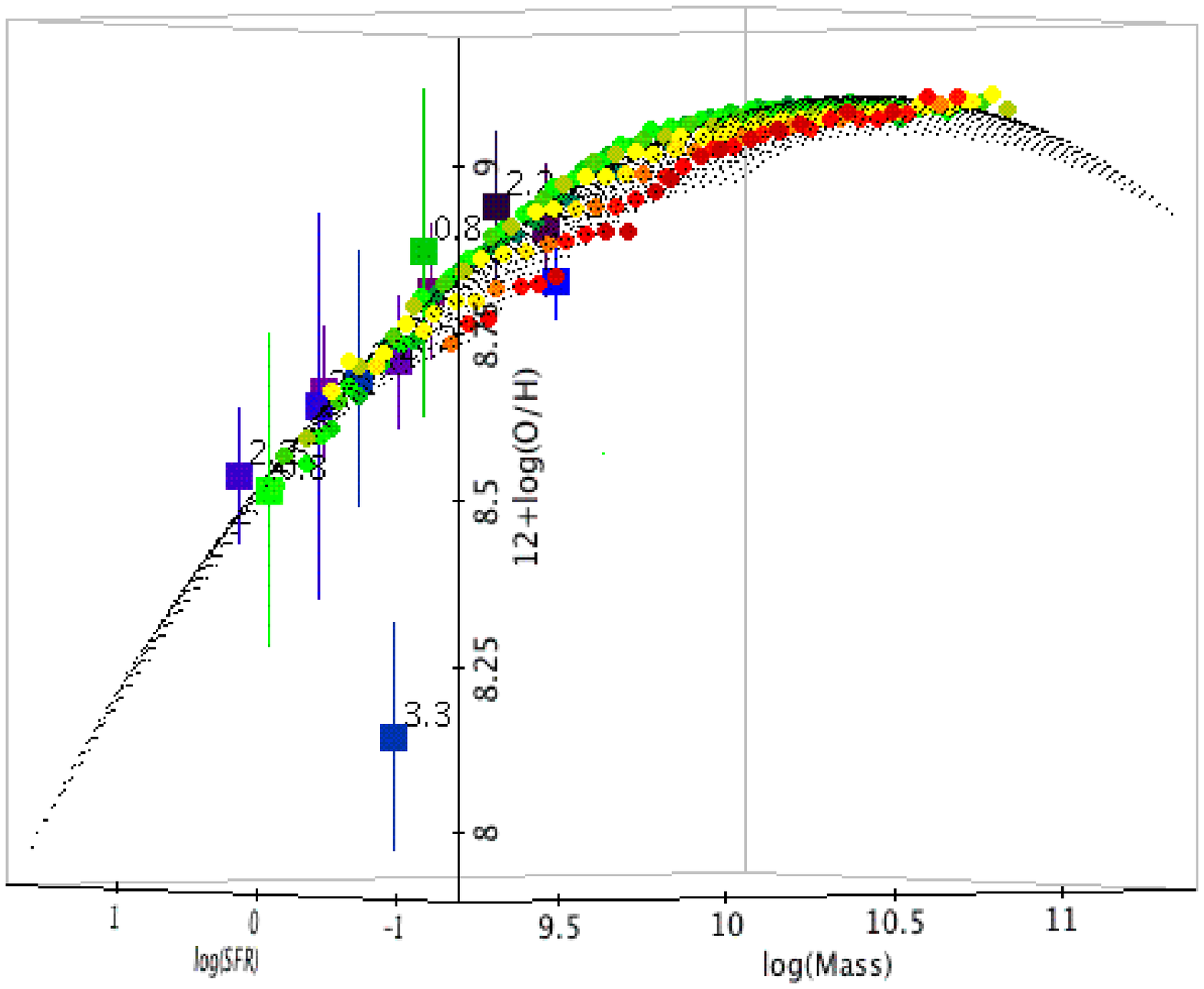} 
	\includegraphics[width=0.32\textwidth]{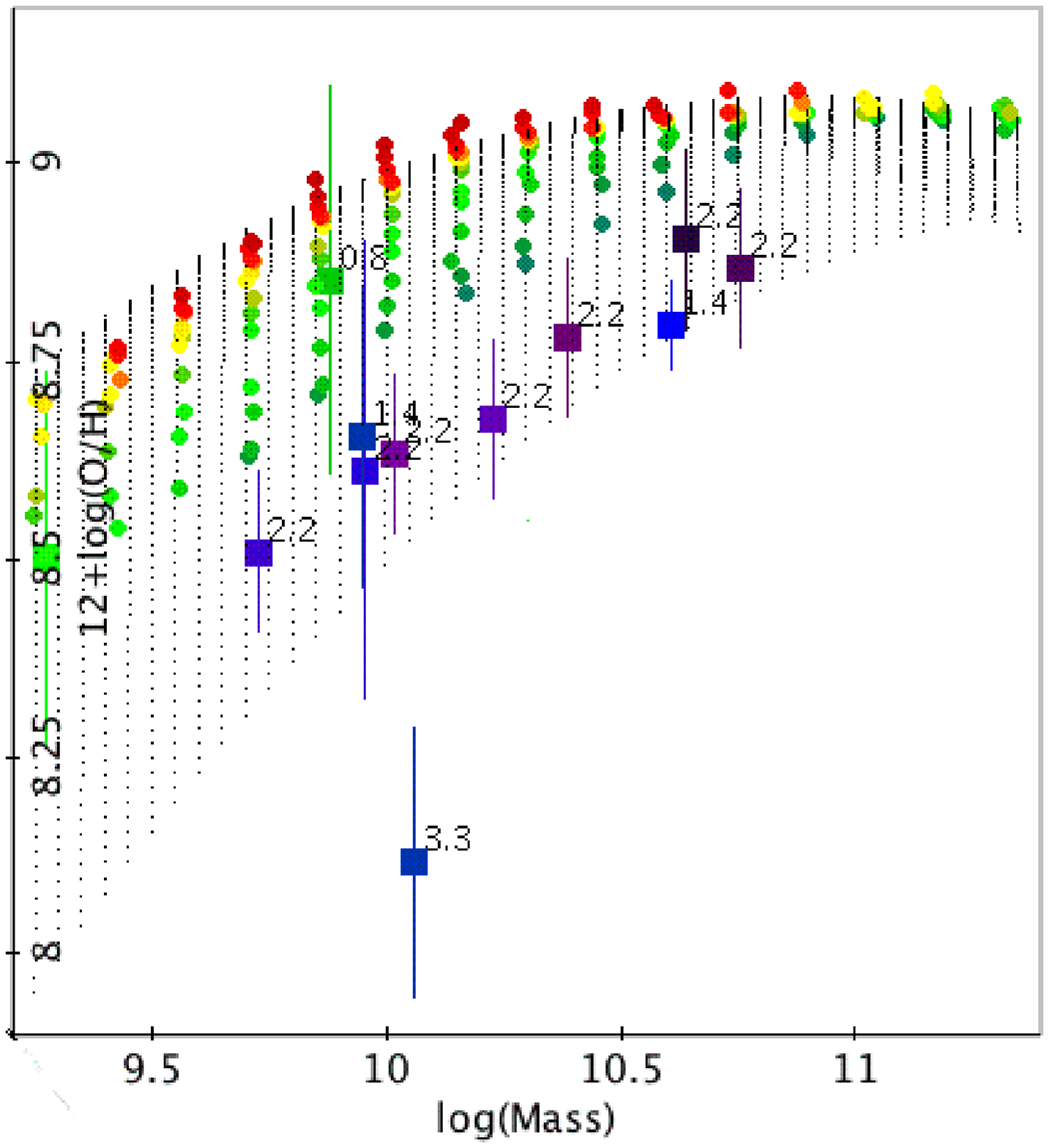} 
}
\caption{\footnotesize
Three projections of the Fundamental Metallicity Relation 
among \mstar, SFR and gas-phase metallicity. 
Circles without error bars are the median values of metallicity of local SDSS galaxies
in bin of \mstar\ and SFR, color-coded with SFR as shown in the colorbar on the right.
These galaxies define a tight surface in the 3D space, with
dispersion of single galaxies around this surface of $\sim$0.05~dex.
The black dots show a second-order fit to these SDSS data, 
extrapolated toward higher SFR.
Square dots with error bars are the median values of high redshift galaxies, 
as explained in the text.
Labels show the corresponding redshifts.
The projection in the lower-left panel emphasizes that most of the high-redshift data,
except the point at z=3.3, are found on the same surface defined by low-redshift data.
The projection in the lower-right panel corresponds to the mass-metallicity relation, as 
in Fig.~\ref{fig:massmet}, showing that the origin of the observed evolution in 
metallicity up to z=2.5 is due to the progressively increasing SFR.
}
\label{fig:cfr1}
\end{figure*}

The dependence of metallicity on \mstar\ and SFR can be better visualized 
in a 3D space with these three coordinates, as shown in Figure~\ref{fig:cfr1}.
SDSS galaxies appear to 
define a tight surface in the space, the Fundamental Metallicity Relation (FMR).
The introduction of the FMR results in a significant reduction of residual 
metallicity scatter with respect to the simple mass-metallicity relation. 
The dispersion of individual SDSS galaxies around the FMR, 
is $\sim$0.06~dex when computed across the full FMR and reduces to $\sim$0.05~dex
i.e, about 12\%, in the central part of the relation where most of the 
galaxies are found.
The final scatter is consistent
with the intrinsic uncertainties in the measure of metallicity 
($\sim$0.03~dex for the calibration, to be added to the uncertainties 
in the line ratios), on mass (estimated to be 0.09~dex by \citealt{Tremonti04}),
and on the SFR, which are dominated by the uncertainties on dust extinction.

The reduction in scatter with respect to the mass-metallicity relation
becomes even more significant when considering
that most of the galaxies in the sample cover 
a small range in SFR, with 64\% of the galaxies ($\pm$1$\sigma$) 
is contained inside 0.8~dex. 
The mass-metallicity relation is not an adequate 
representation of galaxy samples with a larger spread of SFRs, 
as usually find at intermediate redshifts.

\begin{figure*}
\centerline{
	\includegraphics[width=0.48\textwidth]{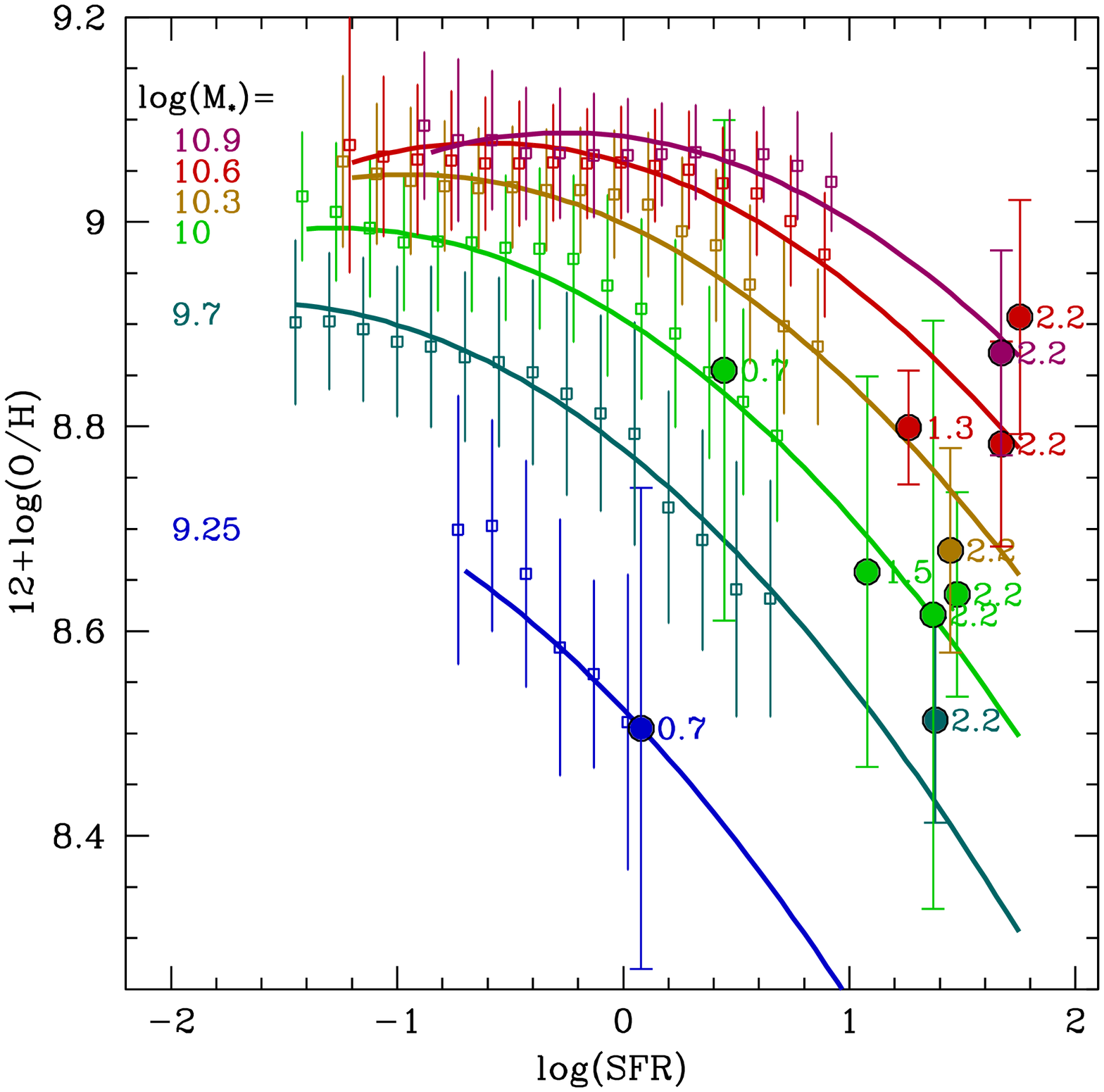} 
	\includegraphics[width=0.48\textwidth]{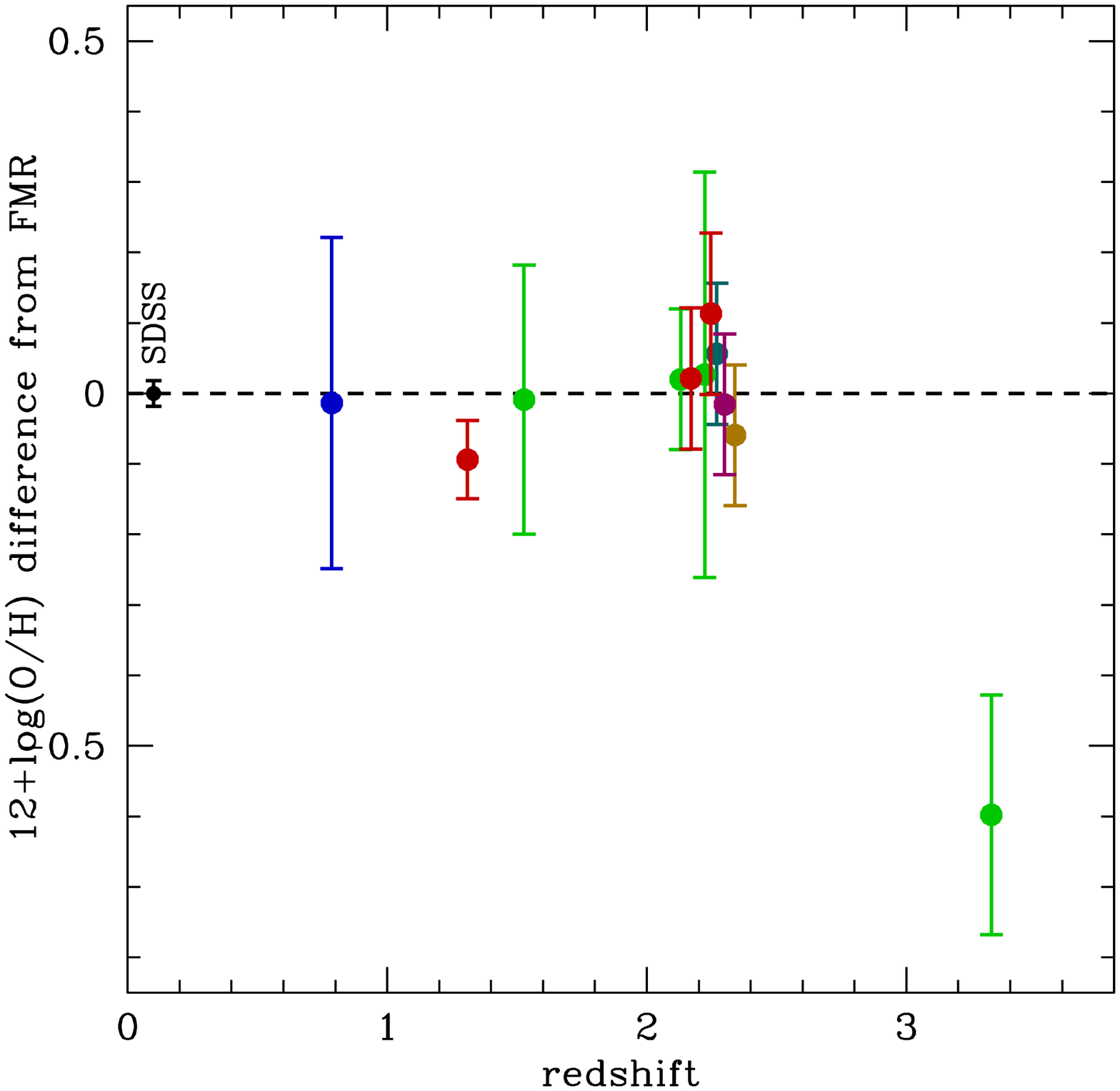} 
}
\caption{\footnotesize
{\em Left:} Metallicity as a function of SFR for galaxies in the three bins 
of \mstar\ containing high-redshift galaxies. The values of log(\mstar) 
are shown by the labels on the left.
Empty square dots are the median values of metallicity of local SDSS galaxies, 
with error bars showing 1$\sigma$ dispersions. 
Lines are the fits to these data.
Solid dots are median values for high-redshift galaxies with z$<$2.5 in the 
same mass bins, with labels showing redshifts.
{\em Right:}
metallicity difference from the FMR for galaxies at different redshifts,
color-coded in mass as in the left panel.
The SDSS galaxies defining the relation are showing at z$\sim$0.1 with their
dispersion around the FMR. All the galaxy samples up to z=2.5 are consistent 
with no evolution of the FMR defined locally. 
Metallicities lower by $\sim$0.6~dex are observed at z$\sim$3.3.
}
\label{fig:plotevol}
\end{figure*}

\section{The FMR at high-redshift}
\label{sec:highz}

The local galaxies can be compared
with several samples of high-redshift objects.
We extracted from the literature samples of galaxies
in four redshift bins, for a total of $\sim$300 objects, 
having published values of emission line fluxes, \mstar, and
dust extinction:
 0.5$<$z$<$0.9 (\citealt{Savaglio05}, GDDS galaxies), 
 1.0$<$z$<$1.6 \citep{Shapley05a,Liu08,Wright09,Epinat09a}, 
 2.0$<$z$<$2.5 \citep{Erb06a,Law09b,Lehnert09,Forster-Schreiber09}, and
 3.0$<$z$<$3.7 \citep{Maiolino08,Mannucci09b}.
The same procedure used for the SDSS galaxies was applied
to these galaxies.

Galaxies at all redshifts follow well defined mass-metallicity relations
(see, for example, \citealt{Mannucci09b}, and references therein).
For this reason each of these samples, 
except the one an z$\sim$3.3 that contains 16 objects only,
is divided into two equally-numerous samples
of low- and high-\mstar\ objects. Median values of \mstar, SFR and 
metallicities are computed for each of these samples.

Galaxies up to z$\sim$2.5 follow the FMR defined locally, with no
sign of evolution.  This is an unexpected result, as simultaneously the
mass-metallicity relation is observed to evolve rapidly with redshift
(see Fig.\ref{fig:massmetevol}).
The solution of this apparent paradox is that distant galaxies have, on average,
larger SFRs, and, therefore, fall in a different part of the same FMR.

In the SDSS sample, metallicity changes more with \mstar\ ($\sim$0.5~dex 
from one extreme to the other at constant SFR, see Fig.~\ref{fig:massmet})
than with SFR ($\sim$0.30~dex at constant mass). 
Therefore mass is the main driver of the level of chemical enrichment 
of SDSS galaxies.
This is related to the fact that 
galaxies with high SFRs, the objects showing the strongest dependence 
of metallicity on SFR (see the right panel of fig.~\ref{fig:massmet}), 
are quite rare in the local universe.
At high redshifts, mainly active galaxies are selected, and the dependence
of metallicity on SFR becomes dominant. 

Galaxies at z$\sim$3.3 show metallicities 
lower of about 0.6~dex with respect to
both the FMR defined by the SDSS sample and galaxies at 0.5$<$z$<$2.5. 
This is an indication that some evolution of the FMR 
appears at z$>$2.5, although its size 
can be affected several potential biases (see \citealt{Mannucci10} for 
a full discussion). A larger data set at z$>$3 is needed to solve this 
question.

\section{What the FMR is telling us}

The interpretation of these results must take into account several effects.
In principle, metallicity is a simple quantity as it is dominated by three processes:
star formation, infall, outflow. If the scaling laws of each of these three processes are
known, the dependence of metallicity on SFR and \mstar\ can be predicted.
In practice, these three processes have a very complex dependence of the properties of the 
galaxies, and can introduce scaling relations in many different ways. 
First, it is not known how {\em outflows}, due to either SNe or AGNs,
depend on the properties of the galaxies.
Second, {\em infalls} of pristine gas are expected to influence metallicity in two ways:
metallicity can be reduced by the direct accretion of metal-poor gas, 
and can be increased by the star formation activity which is likely to follow accretion.
Third, the star formation activity is known to depend on galaxy mass, with
heavier galaxies forming a larger fraction of stars at higher redshifts, and this effect
produce higher metallicities in more massive galaxies.

The dependence of metallicity on SFR can be explained by the dilution effect of the
infalling gas. A simple model can be constructed (see \citealt{Mannucci10})
where a  variable amount of metal-poor, infalling gas, forming stars according to the
Schmidt-Kennicutt law, can explain the dependence of metallicity on SFR.
For this scenario to work, the timescales of chemical enrichment must be longer than
the dynamical scales of the galaxies, over which the SFR is expected to evolve. 
In other words, galaxies on the FMR are in a {\em transient phase}:
after an infall, galaxies first evolve towards higher SFR 
and lower metallicities. Later, while gas is converted into stars and new metals are produced, 
either galaxies drop out of the sample because they have faint \ha, or evolve toward higher 
values of mass and metallicities along the FMR. 
In this scenario, the dependence of metallicity on SFR is due to infall and dominates at 
high redshifts, where galaxies with massive infalls and large SFRs are found. In contrast,
in the local universe such galaxies are rare, most of the galaxies have low level of accretion,
and abundances are dominated by the
dependence on mass, possibly due to outflow.

In many local galaxies, timescales of chemical enrichment can be shorter 
than the other relevant timescales (e.g., \citealt{Silk93}), 
and galaxies can be in a {\em quasi steady-state situation},
in which gas infall, star formation and metal ejection occur simultaneously
\citep{Bouche09}.
Assuming this quasi steady-state situation, in which infall and SFR are 
slowly evolving with respect to the timescale of chemical enrichment, 
it can be shown \citep{Mannucci10}
that our results support a scenario where outflows are
inversely proportional to mass and increase with SFR$^{0.65}$.\\

The small scatter of SDSS galaxies around the FMR
can be used to constrain the characteristics of gas accretion.
For this infall/outflow scenario to work and produce a very small scatter 
round the FMR, two conditions are simultaneously required:
(1) star formation is always associated to the same level of metallicity
dilution due to infall of metal-poor gas; 
(2) there is a relation between the amount of infalling and outflowing gas and 
the level of star formation.
These conditions for the existence of the 
FMR fits into the smooth accretion models proposed by several groups
\citep{Bournaud09,Dekel09}, where continuos infall of pristine gas is the main
driver of the grow of galaxies. In this case, metal-poor gas is continuously accreted
by galaxies and converted in stars, and a long-lasting equilibrium between gas accretion, 
star formation, and metal ejection is expected to established.

\begin{figure*}[t]
\label{figmet}
\centerline{\includegraphics[width=0.8\textwidth]{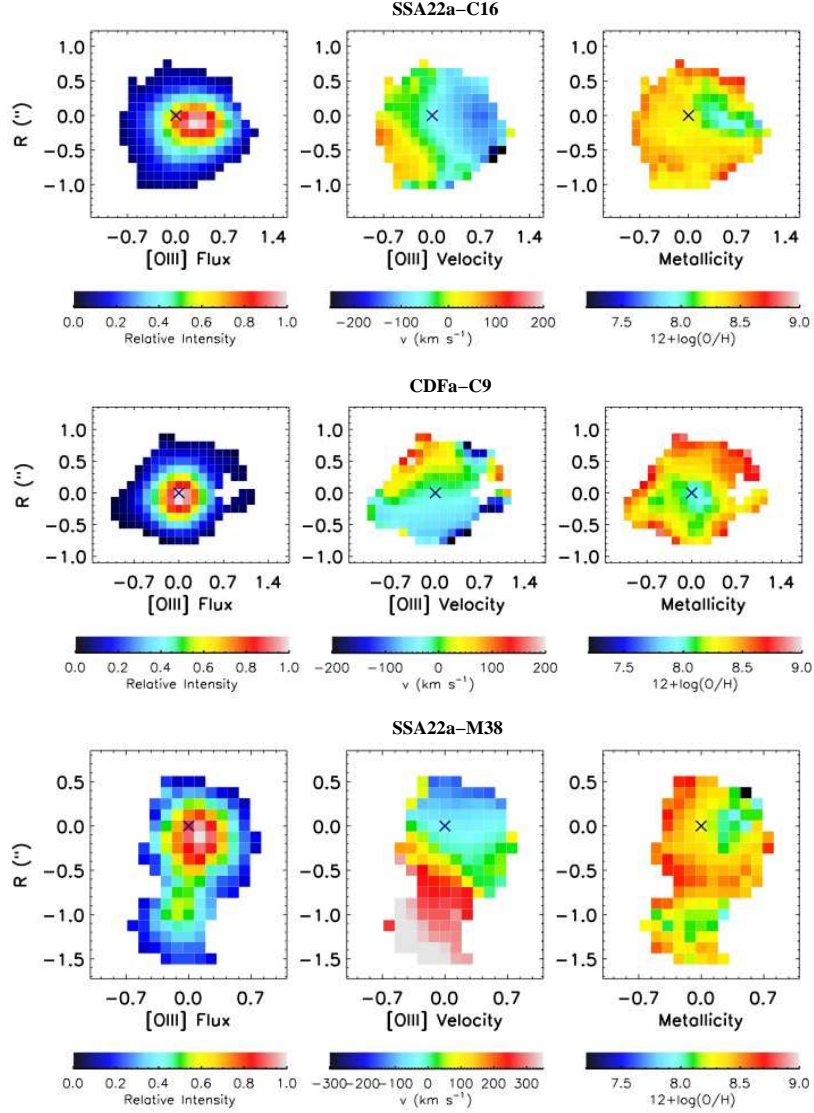}}
\caption{\footnotesize
Surface brightness of the [OIII]$\lambda$5007 line, velocity map, 
and gas phase metallicity, plotted as relative abundances of oxygen and 
hydrogen parameterized in units of $12+log(O/H)$, of the three 
galaxies in \cite{Cresci10}. 
Lower metallicity region are surrounded by a more enriched disk. 
The crosses in each panel mark the position of the continuum peak.}
\end{figure*}

\section{Abundance gradients in high-redshift galaxies}

Recently \citep{Cresci10}, we have obtained a direct evidence of the presence of smooth 
accretion of gas in high redshift galaxies. 

We selected three Lyman-break galaxies  among the AMAZE \citep{Maiolino08} 
and LSD \citep{Mannucci09b} samples which show a remarkably symmetric velocity field in 
the [OIII] emission line, which traces the ionized gas kinematics 
(see Fig. \ref{figmet}). Such kinematics indicates that these are rotationally supported 
disks (Gnerucci et al., in preparation), 
with no evidence for more complex merger-induced dynamics.
Near-infrared spectroscopic observations of the galaxies were obtained with the integral 
field spectrometer SINFONI on VLT,
and we used the flux ratios between the main rest-frame optical lines to obtain the metallicity map
shown in Fig. \ref{figmet}. An unresolved region with lower metallicity 
is evident in each map, surrounded by a more uniform disk of higher metal content. In one case, CDFa-C9, 
the lower metallicity region is coincident with the galaxy center, as traced by the continuum peak, 
while it is offset by $\sim 0.60''$ (4.6 kpc) in SS22a-C16 and $\sim0.45''$ (3.4 kpc) in SS22a-M38. 
On the other hand, in all the galaxies the area of lower metallicity is coincident or closer than $0.25''$ 
(1.9 kpc, half of the PSF FWHM) to the regions of enhanced line emission, 
tracing the more active star forming regions.
The average difference between high and low metallicity regions 
in the three galaxies is  $0.55$ in units of 12+log(O/H), 
larger than the $\sim0.2-0.4$ dex gradients measured in the Milky Way and other 
local spirals \citep{van-Zee98} 
on the same spatial scales. The measured gas phase abundance variations have a 
significance between 98\% and 99.8\% .
It can be shown \citep{Cresci10} that variations of 
ionization parameter across the galaxies cannot explain the observed gradients of
 line ratios, and that different metallicities are really requested.

Current models of chemical enrichment in galaxies \citep{Molla97} cannot reproduce our observations 
at the moment, as they assume radially isotropic gas accretion onto the disk and the instantaneous recycling 
approximation. Nevertheless, the detected gradients can be explained in the framework of the 
cold gas accretion scenario  \citep{Keres05}  recently proposed to explain the properties of gas rich, rotationally 
supported galaxies observed at high redshift \citep{Cresci09,Forster-Schreiber09}. In this scenario, the observed low 
metallicity regions are created by the local accretion of metal-poor gas in clumpy streams \citep{Dekel09}, 
penetrating deep onto the galaxy following the potential well, and sustaining the observed high star formation 
rate in the pre-enriched disk. Stream-driven turbulence is then responsible for the fragmentation of the disks 
into giant clumps, as observed at $z \geq 2$ \citep{Genzel08,Mannucci09b}, that are the sites of 
efficient star formation and possibly the progenitors of the central spheroid. This scenario is also in 
agreement with the dynamical properties of our sample, which appears to be dominated by gas rotation in 
a disk with no evidence of the dynamical asymmetries typically induced by mergers.
The study of the relations between metallicity gas fractions, effective yields, and SFR \citep{Cresci10}
show that the low-metallicity regions can be well explained by amounts of infalling gas
much larger than in the remaining high-metallicity regions.

Our observations of low metallicity regions in these three galaxies at $z\sim3$ 
therefore provide the evidence for the actual presence of accretion of metal-poor gas 
in massive high-z galaxies, capable to sustain high star formation rates without 
frequent mergers of already evolved and enriched sub-units. 
This picture was already indirectly suggested by recent observational studies of 
gas rich disks at $z\sim1-2$ \citep{Forster-Schreiber09,Tacconi10}, 
and is in agreement with the FMR describe above.

\bibliographystyle{/Users/filippo/arcetri/Papers/aa-package/bibtex/aa}
\bibliography{/Users/filippo/arcetri/bibdesk/Bibliography}

\end{document}